\documentclass[journal]{IEEEtran}

\usepackage{flushend}  
\usepackage{etoolbox}  

\newtoggle{Anonymous}
\toggletrue{Anonymous}
\togglefalse{Anonymous}

\makeatletter
\usepackage{blindtext}
\usepackage{eso-pic}
\usepackage{cite}
\usepackage{amsmath,amssymb,amsfonts}
\usepackage{algorithmic}
\usepackage{graphicx}
\usepackage{textcomp}
\usepackage{xcolor}
\def\BibTeX{{\rm B\kern-.05em{\sc i\kern-.025em b}\kern-.08em
    T\kern-.1667em\lower.7ex\hbox{E}\kern-.125emX}}
    
\usepackage{eso-pic}

\newcommand{\E}{\mathrm{e}} 
\newcommand{\I}{\mathrm{i}} 
\newcommand{\D}{\mathrm{d}} 
\newcommand{\DD}{\mathrm{D}} 

\newcommand{\J}{\mathrm{i}}
\renewcommand{\epsilon}{\varepsilon}
\renewcommand{\vec}{}

\begin{document}
\title{\vspace*{1cm}Natural Occurrence of \\Fractional Derivatives in Physics}


\iftoggle{Anonymous}
{
\author{\IEEEauthorblockN{} 
\IEEEauthorblockA{\textit{ }\\
  \\
  \\}
}
}
{
\author{\IEEEauthorblockN{Sverre Holm}
\IEEEauthorblockA{\textit{Department of Physics, University of Oslo}\\
Oslo, Norway \\
ORCID https://orcid.org/0000-0002-2273-705X}
}
}

\maketitle

\begin{abstract}
Power laws in time and frequency appear in fields such as linear viscoelasticity and acoustics, viscous boundary layer problems, and dielectrics. This is consistent with fractional derivatives in the fundamental descriptions, since power laws in time and frequency are related by the Fourier transform, and also associated with fractional derivatives. Examples here include power-law frequency dependent attenuation in ultrasound, elastography and sediment acoustics. In viscous boundary problems there is a viscodynamic operator in the Biot poroviscoelastic theory which may be formulated with a fractional derivative. Power law and stretched exponential temporal responses of non-ideal capacitors can also be shown to relate to the Cole-Cole power-law dielectric model.
\end{abstract}
\begin{IEEEkeywords}
power law, stretched exponential, Biot theory, Abel, Cole-Cole model
\end{IEEEkeywords}

\section{Introduction}
\label{Sect:Introduction}
Temporal power-law responses and frequency domain power laws appear in many fields of physics, both in empirical results and in physical theories.
Here we outline such naturally occurring power laws in three fields: 
\begin{itemize}
\item linear viscoelasticity and acoustics
\item viscous boundary layer problems
\item dielectrics
\end{itemize}
In Fourier analysis a power law in frequency is  transformed into a power law in  time:
\begin{equation}
f(t) = \frac{t^{-\beta}}{\mathrm{\Gamma}(1-\beta)}, \enspace t>0 \quad \leftrightarrow \quad F(\omega) = (\I \omega)^{\beta-1},
\label{eq:FourierPower}
\end{equation}
where $\Gamma(x)$ is the gamma function, and $\Gamma(n) = (n-1)!$. This result demonstrates that  power laws in time and frequency are closely related. 
%
Further, a fractional or non-integer derivative of the Caputo and Riemann-Liouville kinds can be considered as a generalization of the differentiation property of the Fourier transform:
\begin{equation}
\frac{\D^\alpha}{\D t^\alpha} f(t) \quad \leftrightarrow \quad  (\I \omega)^\alpha F(\omega)
\label{eq:FourierDefFractional}
\end{equation}
This shows that power laws  are also associated with fractional derivatives. 

Here we explore how  common power-law descriptions in many field relate to temporal descriptions with fractional derivatives.


\section{Observations of power laws}
\subsection{Linear viscoelasticity and acoustics}
\label{Sec:LinearViscoleasticity}

In 1921, Nutting showed that the relaxation modulus, the stress response to a unit step in strain, for complex materials such as paint and oil, deviates from a simple exponential and instead varies according to a power law, \cite{nutting1921new}:
\begin{equation}
G(t) \propto t^{-\alpha}, \quad 0<\alpha<1.
\label{eq:NuttingRelaxation}
\end{equation}
It is now acknowledged that this property applies to many biological and non-biological materials. Scott-Blair was also the first to suggest that non-integer, i.e.~fractional derivatives, could describe this property, \cite{blair1951rheological}.


In all  media which can be modeled as a linear time-invariant system, the attenuation as a function of distance, $x$, for a sinusoidal input  of frequency $\omega$, is:
\begin{equation}
|u(x)| \propto \E^{-\alpha_k(\omega) x},
\label{eq:GeneralAttenuation}
\end{equation}
where $\vec{u}$ can be particle velocity or pressure variation. Often the attenuation follows a power law as a function of frequency: 
\begin{equation}
\alpha_k(\omega) = \alpha_0 \omega^y,
\label{eq:PowerLawAttenuation}
\end{equation}
In many fields, the exponent, $y$, is often close to unity. This was shown by measurements in blood in the 0.8-3 MHz range early in the 1950's \cite{carstensen1953determination}, and confirmed in numerous experiments since then \cite{parker2022power}. This property  is valid both for compressional and shear waves over many decades of frequency and extends to many other materials also. The review in\cite{Szabo00} cites e.g.~references back to 1965 for  attenuation that varies linearly with frequency  in seismics.

In many fields, including seismics, this kind of attenuation is usually  called constant-Q attenuation. This can be seen from the definition of the quality factor or $Q$-value, which usually is given in the form of its inverse:
\begin{equation}
Q^{-1}(\omega) = \frac{2 \alpha_k(\omega) c_{ph}(\omega)}{\omega} 
\label{eq:approxQ}
\end{equation}
It can be interpreted as being proportional to the penetration depth of the wave, measured in number of wavelengths 
\cite[2.3]{holm2019waves}.
An attenuation which increases linearly with frequency, $y=1$ in \eqref{eq:PowerLawAttenuation}, is associated with a constant phase velocity, $c_{ph}(\omega)$, and is therefore constant-$Q$ propagation.

\subsection{Viscous boundary layers}
\label{Sec:ViscousBoundaryLayer}

The friction when a fluid passes over a surface due to wave motion occurs in a viscous boundary layer with thickness:
\begin{equation}
\delta_{BL} =\sqrt{\frac{2 \eta}{\rho \omega}}
\label{eq:boundaryLayer}
\end{equation}
where $\omega$ is frequency, $\rho$ is fluid density, and $\eta$ is the viscosity \cite[Sect.~7.6]{holm2019waves}. This is parallel to the skin depth for electrical current in a conductor, where the effective depth of penetration follows a similar inverse square root dependency of frequency. 

In a tube, the smaller the thickness of the boundary layer relative to the radius, the more the flow will be affected. This becomes evident in a  poroviscoelastic medium which consists of fluid filled pores in a solid, and which is a  model for e.g.~porous rocks and bones. Biot theory describes waves in such media and contains a correction factor due to wave motion in the pores 
 \cite{biot1956theoryII}. The factor is called a viscodynamic operator or drag coefficient. It is unity at low frequencies where the viscous boundary layer of \eqref{eq:boundaryLayer} is large compared to the pore dimension. As frequency increases, the thickness of the boundary layer decreases, and the viscodynamic operator increases. 
The correction factor can  be interpreted as the increase in effective viscosity with frequency \cite[Sect.~7.6]{holm2019waves}.

In a circular pore, the  viscodynamic operator is:
\begin{equation}
{F}(z) =   \frac{z}{4}\frac{\mathrm{I}_1(z)}{\mathrm{I}_2(z)}, \quad \I z= \I^{3/2} \sqrt{\omega \tau}, \quad \tau = \frac{\rho R^2}{\eta},
\label{eq:ViscodynamicOperator}
\end{equation}
where $\mathrm{I}_\nu$ is the modified Bessel function of the first kind of order $\nu$, and $R$ is pore radius. The product $\omega \tau$ determines which regime the flow operates in. If it is much smaller than unity, the flow is laminar (Poiseuille flow), and if it is large, the flow is turbulent.

Biot showed that  the viscodynamic operator of \eqref{eq:ViscodynamicOperator} can be approximated as a square root of frequency  for large argument.
More accurately, it can also be approximated as  \cite{chandrasekaran2022wave}:
\begin{equation}
{F}(\omega) \approx \sqrt{1+ \I \frac{\omega \tau}{16}}
\approx  \frac{1}{4} \sqrt{\I \omega \tau}
= \frac{R}{\delta_{BL}} \sqrt{\frac{\I}{8}},
\label{eq:Viscodynamic}
\end{equation}
thus increasing with the ratio of radius and the boundary layer thickness for large arguments. This result is robust to changes in pore geometry as a similar square root dependency of frequency is found even for a 2D layered pore \cite{biot1956theoryII}.

\subsection{Observations in dielectrics}
\label{Sec:Dielectrics}

The current step response of an ideal capacitor in series with a resistor is an exponential, but for a non-ideal capacitor it may be a power law instead. This is the Curie-von Schweidler law with roots  back to 1889 and 1907:
\begin{equation}
I(t) \propto t^{\alpha-1}.
\label{eq:CurieVonSchweidler}
\end{equation}

Another classical result is that the discharge of a capacitor in a Leiden jar may follow a stretched exponential:
\begin{equation}
 Q(t) \propto \exp{\left[-(t/\tau)^\alpha\right]},
 \label{eq:KWW}
 \end{equation} 
%
where $Q$ means charge (i.e.~a different meaning than in \eqref{eq:approxQ}),  $\tau$ is a time constant, and  $\alpha$ is order. This result dates back to 1854
\cite{kohlrausch1847ueber} and is often called the Kohlrausch-Williams-Watt model \cite{williams1970non}.

In the 1950's it was found that in complex materials such as liquids and polymers, the permittivity follows \cite{Cole1941}:
\begin{equation}
{\epsilon_r}(\omega) 
= \epsilon_\infty + \frac{\epsilon_s - \epsilon_\infty}{1 +( \I \omega \tau)^{\alpha}}, \quad 0<\alpha \le 1,
\label{eq:ColeCole}
\end{equation}
where $\epsilon_s$ iand $\epsilon_\infty < \epsilon_s$ are the limits at 0 and infinite frequencies respectively. 
In the field of electromagnetics, the exponent is usually $1-\alpha'$ instead of $\alpha$, but here the latter terminology is used instead in order to harmonize it with the other fields discussed here. The Cole-Cole model is a generalization of Debye relaxation ($\alpha=1$).


\section{Power laws and fractional calculus}
In all the examples of the previous section, power laws in either time or frequency appear. It will now be shown that this is closely linked to non-integer derivatives. This is done by starting with the  definition of the Fourier transform: 
\begin{equation}
F(\omega) = \mathcal{F}\left( f(t) \right) = \int_{\infty}^\infty  f(t) \, \E^{-\I \omega t} \; \D t.
\end{equation}
A few functions will easily generalize to a fractional derivative. Among them is the exponential where direct extension of the formula for the integer order derivative gives:
\begin{equation}
\DD_t^{\alpha}\E^{k t} =  \frac{\D^\alpha}{\D t^\alpha} \E^{k t} = k^\alpha \E^{k t}, \quad k \ge 0,
\end{equation}
where $\DD_t^{\alpha}$ is a compact description of a fractional derivative which usually is written  $_a \DD_t^{\alpha}$ where $a$ is the lower limit in the defining integral. Here we always assume  that $a=-\infty$ as in \eqref{eq:Caputo} and \eqref{eq:RL}. Because $k$ can also be complex, the derivative of the exponential enables a generalized 
differentiation property for the Fourier transform:
\begin{equation}
\DD_t^\alpha f(t) = \frac{1}{2 \pi}  \int_{\infty}^\infty  F(\omega)  \DD_t^\alpha  \E^{\I \omega t} \; \D \omega 
\end{equation}
which leads to the property in \eqref{eq:FourierDefFractional} and also repeated below.
Eq.~\eqref{eq:FourierDefFractional}  is the simplest definition of a fractional derivative.

The time domain interpretation of the fractional derivative is  more complex and can be found by splitting the order $\alpha$  into an integer part, $m = \lceil \alpha \rceil$, and a remainder, $\alpha-m \le 0$:
\begin{equation}
\mathcal{F} \left( \DD_t^\alpha f(t) \right) = (i\omega)^{\alpha} F(\omega)
=\left[ (i\omega)^{m} F(\omega)\right] \cdot  (i\omega)^{\alpha-m}.
\label{eq:FractionalProduct}
\end{equation}
The first part, contained in the brackets, corresponds to an ordinary derivative of order $m$. But this is an order which is higher than the desired one, so the remainder, $(i\omega)^{\alpha-m}$ is equivalent to an integration of order $m-\alpha$.

By letting $\beta=\alpha+1-m$ in  \eqref{eq:FourierPower}, it is seen that the product in \eqref{eq:FractionalProduct}  corresponds to convolution with a temporal power law:
\begin{equation}
\DD_t^\alpha f(t)
= \DD_t^m f(t) * \frac{1}{\Gamma(m-\alpha)}\frac{1}{t^{\alpha+1-m}}. 
\end{equation}
The differentiation property of convolution follows from the commutativity of convolution:
\begin{equation}
\left( \frac{\D}{\D t}f(t) \right) \ast g(t)=
\frac{\D}{\D t} \left( f(t) \ast g(t) \right).
\label{eq:ConvolutionDifferentiation}
\end{equation}
This property gives rise to two different definitions of the fractional derivative. If one first performs integer order differentiation and then convolution according to the left-hand side of \eqref{eq:ConvolutionDifferentiation}, one gets the Caputo definition:
\begin{equation}
^C D_t^\alpha f(t) = \frac{1}{\Gamma(m-\alpha)} \int_{-\infty}^t \frac{\DD_\tau^m f(\tau)}{(t-\tau)^{\alpha+1-m}}  \;  \D\tau.
\label{eq:Caputo}
\end{equation}
%
The first mention of fractional differentiation was  by Niels Henrik Abel in one of his first articles \cite{abel1823oplosning}, using $0<\alpha<1$ and $m=1$ above. The fractional derivative was unfortunately not pursued in Abel's later work \cite{podlubny2017niels}.

On the other  hand, if a  convolution is first performed and then the integer order differentiation, as on the right-hand side of \eqref{eq:ConvolutionDifferentiation}, one gets the Riemann-Liouville definition:
\begin{equation}
^{RL} D_t^\alpha f(t) =\frac{1}{\Gamma(m-\alpha)}  \DD_t^m   \int_{-\infty}^t \frac{f(\tau)}{(t-\tau)^{\alpha+1-m}}  \; \D\tau.
\label{eq:RL}
\end{equation}

This derivation of a fractional derivative from the frequency domain demonstrates these important properties:
\begin{itemize}
\item Power laws in time and frequency are equivalent
\item A power laws is equivalent to a fractional derivative
\end{itemize}
These properties are the key to formulating the medium characteristics of Sect.~\ref{Sect:Introduction} in terms of fractional derivatives.

\section{Linear viscoelasticity and acoustics}
The observations noted in Sect.~\ref{Sec:LinearViscoleasticity} can be explained by considering fractional constitutive laws.
For reference, if one combines linearized conservation of linear momentum, linearized conservation of mass, with the constitutive law of a viscous medium, one gets the classical viscous wave equation:
\begin{align}
\nabla^2 \vec{u}  - \frac{1}{c_0^2} \frac{\partial^2 \vec{u}}{\partial t^2} + \frac{\eta}{E} \frac{\partial}{\partial t} \nabla^2 \vec{u} = 0, \quad c_0^2 = \frac{E}{\rho},
\label{eq:KelvinVoigtWave}
\end{align}
where $c_0$ is speed of sound in the limit as  frequency approaches zero. The equation corresponds to a medium model with a spring of elasticity, $E$, in parallel with a Newtonian viscosity $\eta$ and can also be found from the Navier-Stokes equation. The underlying constitutive law between stress, $\sigma(t)$, and strain, $\epsilon(t)$,  in this case is
\begin{equation}
\sigma(t) =  E \epsilon(t) + \eta \frac{\partial \epsilon(t)}{\partial t}.
\label{eq:Constitutive2Ch3}
\end{equation}
The losses are represented by the viscosity and the attenuation, $\alpha_k$ for small losses, $\omega \eta/E \ll 1$, is quadratic in frequency, i.e. $y=2$ in \eqref{eq:PowerLawAttenuation}.

In order to model media with other values of exponent, $y$, the wave equation can be made fractional. This can be done in several different ways, e.g.~both by changing the Laplacian in the third term to a fractional one or by changing the temporal derivative. It is only the latter which easily can be interpreted as a realistic medium model  consisting of a spring in parallel with a spring-pot, a fractional viscosity. This is the fractional Kelvin-Voigt model:
\begin{equation}
\sigma(t)  =  E \epsilon(t) + \eta\frac{\partial^{\alpha}\epsilon(t)}{\partial t^{\alpha}}.
\label{eq:KelvinVoigtFractional}
\end{equation}
This leads to the fractional Kelvin-Voigt wave equation
\begin{align}
\nabla^2 \vec{u}  - \frac{1}{c_0^2} \frac{\partial^2 \vec{u}}{\partial t^2} + \frac{\eta}{E} \frac{\partial^\alpha}{\partial t^\alpha} \nabla^2 \vec{u} = 0.
\label{eq:FractionalKelvinVoigtWave}
\end{align}
It can be shown that its attenuation will increase with power $y=\alpha+1$. 
When such a model is fitted to measurements from Magnetic Resonance Elastography (MRE) of low-frequency (30-100 Hz) shear waves in human tissue,  the parameters have been shown to be able to discriminate  between inflammation and fibrosis in the liver \cite{sinkus2018rheological}. This is also the model for compressional waves in one of the simpler models for sub-bottom sediment, the Grain Shearing model \cite{pandey2016connecting}. 

The relaxation response of the fractional Kelvin-Voigt model is a power law:
\begin{equation}
G(t) =   E + \eta \; \frac{t^{-\alpha}}{\mathrm{\Gamma}(1-\alpha)}.
\label{eq:RelaxationFractionalKV}
\end{equation}
Here one can recognize the relaxation response first proposed by Nutting in \eqref{eq:NuttingRelaxation}, especially if the spring is removed, $E=0$. Thus the fractional Kelvin-Voigt model may explain both the power-law relaxation response and the power-law attenuation observed in many media.

Removal of the spring, also leads to an even simpler wave equation, the fractional diffusion-wave equation, analyzed in \cite[Chaps.~6-7]{Mainardi2022}:
\begin{align}
\nabla^2 \vec{u} - \frac{\rho_0}{\eta} \frac{\partial ^{2-\alpha}}{\partial t^{2-\alpha}} \vec{u}= 0.
\label{eq:FractionalDiffusionWave2}
\end{align}
It also results in near linear variation of attenuation with frequency as $\alpha \rightarrow 0$. Often this simplified model which only builds on a spring-pot medium model \cite{Meral2010}, is adequate for describing shear waves in human tissue in elastography \cite{parker2019towards}.  
It also models shear waves in the Grain Shearing model \cite{pandey2016connecting}.

Power laws are also applicable to describe the bio-mechanical properties of cells and the connection with fractional derivatives in that field was  recognized early \cite{djordjevic2003fractional}.


\subsection{Glassy models}
Simple media like seawater and air have an attenuation which is  modeled as a sum of three relaxation terms.
%
Properties of more complex media are usually explained as the result of multiple such relaxation processes over a large spread of relaxation frequencies:
\begin{equation}
\alpha_k(\omega) = \omega^2 \int_0^\infty A(\Omega) \frac{ \Omega}{\omega^2+\Omega^2} \D \Omega,
\label{eq:MultipleRelaxation}
\end{equation}
where $\Omega$ is relaxation frequency. A particular set of relaxation strengths $A(\Omega)$ may result in the power-law attenuation of \eqref{eq:PowerLawAttenuation}.

There is evidence that for instance soft cells may behave as a strong glass under influence of shear  \cite{zhou2009universal}. In glasses it is common to adopt a more physical point of view and consider the energy rather than the frequency of the relaxation processes. Attenuation as a function of absolute temperature, $T$, and frequency in a multiple relaxation material is then \cite{carini2005ultrasonic}:
\begin{equation}
\alpha_k (T, \omega) \propto \frac{\omega^2}{T} \int_0^\infty g(E) \frac{ \tau(E)}{1+\omega^2 \tau^2(E)} \D E, \enspace \tau =1/\Omega,
\label{eq:MultipleRelaxationEnergy}
\end{equation}
where $g(E)$ is an energy distribution and $\tau$ is relaxation time which may relate to thermal activation energies.
%
%
%
An example of an energy distribution function in glass is a Gaussian \cite{carini2005ultrasonic}. 
This point of view may give new insight in how power-law attenuation relates to physical parameters.

\section{Viscous boundary layers}
One of the first to note that the boundary layer properties of Sect.~\ref{Sec:ViscousBoundaryLayer} relate to fractional derivatives was \cite{Torvik1984}. They considered Stokes' second problem which deals with a uniformly oscillating plate on the surface of a viscous fluid.
%

The approximation of the operator of \eqref{eq:Viscodynamic} to inverse square dependency of frequency, \eqref{eq:FourierDefFractional}, shows that it corresponds to a temporal operator with a half order fractional derivative. When applied to an arbitrary input, $x(t)$, the result is:
\begin{equation}
f(t) \ast x(t)
= \frac{\sqrt{\tau}}{4} \DD_t^{1/2} x(t).
\label{eq:viscodynamicHighFrequency}
\end{equation}
This result was given in \cite{enelund2010time} and enables the finding of the high-frequency asymptote of the wave equation from the rather complicated Biot dispersion equations \cite[Sect.~8.4]{holm2019waves}. This result can be extended to all frequencies if the operator corresponding to the first approximation of \eqref{eq:Viscodynamic} is found instead. Consider this frequency domain factor:
\begin{equation}
    F(\omega) = \left( 1 + \I \omega/\Omega_0 \right)^\gamma, \quad \Omega_0 = 1/\tau.
    \label{eqn:CD-factor}
\end{equation}
where we are mainly interested in the case for $\gamma=0.5$. This factor corresponds to:
   \begin{align}
        f(t) \ast x(t) & =  \E^{-\Omega_0 t} \DD_t^\gamma \left( \E^{\Omega_0 t} x(t) \right)\\ \notag
        &=\left(\DD_t + \Omega_0 \right)^{\gamma} x(t).
        \label{eq:FractionalPseudo}
    \end{align}
The latter expression is called  the  fractional pseudo-differential operator, \cite{nigmatullin1997cole}, also found in \cite[App.~B.2]{garrappa2016models}. 


The expression of the viscodynamic operator in terms of the  fractional pseudo-differential operator enables the finding of wave equations for all three wave modes of the Biot theory, the fast and slow compressional modes and the shear mode. These wave equations describe the entire frequency range, from low frequencies where the viscodynamic operator is unity, and into the high-frequency turbulence region  \cite{chandrasekaran2022wave}.

\section{Dielectrics}
There are also connections between the empirical time domain and frequency results of Sect.~\ref{Sec:Dielectrics}.
The relative permittivity of \eqref{eq:ColeCole} is one of several models for how the displacement field, D relates to the electric field E: 
\begin{equation}
D = \varepsilon_0 \varepsilon_r E.
\label{eq:D-E}
\end{equation}
If $ \varepsilon_r$ is a constant the dielectric gives rise to capacitance:
\begin{equation}
C = \frac{A }{d} \varepsilon_0 \varepsilon_r,
\end{equation}
where $A$ is area and $d$ is distance between plates. 

The simplest medium model is the Debye relaxation model from 1912 which describes liquids and solids. Since $\alpha=1$ in \eqref{eq:ColeCole} in that case, the only frequency dependent factor is $\I \omega$, corresponding to ordinary differentiation. Putting  \eqref{eq:ColeCole} in \eqref{eq:D-E} in that case, implies that the time domain equivalent of relaxation is:
\begin{equation}
D(t) + \tau \frac{\D D(t)}{\D t} = \varepsilon_0\varepsilon_s E(t) + \tau  \varepsilon_0\varepsilon_\infty \frac{\D E(t)}{\D t}.
\label{eq:DebyeTime}
\end{equation}

The Cole-Cole model with arbitrary $\alpha$ is used in fields as diverse as bioimpedance and electrochemistry. By letting $\varepsilon_\infty \rightarrow 0$ and $\enspace (\omega \tau)^\alpha \gg 1$ in \eqref{eq:ColeCole}, the model simplifies to $ {\epsilon_s}/{( \I \omega \tau)^{\alpha}}$. This corresponds to a generalized capacitor:
\begin{equation}
C \approx \frac{A}{d}  \frac{\varepsilon_0 \varepsilon_s}{(\I \omega \tau)^{\alpha}}.
\end{equation}
The impedance is proportional to $(\I \omega \tau)^{\alpha-1}$ and can have any phase angle between voltage and current from a resistor's 0\textdegree{} to an ordinary capacitor's -90\textdegree{}. As it is independent of frequency, it is called a Constant Phase Element (CPE). It also exists in a less common inductive version where impedance is proportional to $(\I \omega \tau)^{1-\alpha}$ and the phase angle varies from 0\textdegree{}  to  90\textdegree{}\cite{holm2021simple}.

The  Cole-Cole medium model gives rise to the following time domain equation of state which generalizes \eqref{eq:DebyeTime}:
\begin{equation}
D(t) + \tau^\alpha \frac{\D^\alpha D(t)}{\D t^\alpha} = \varepsilon_0 \varepsilon_s E(t) + \tau^\alpha \varepsilon_0 \varepsilon_\infty \frac{\D^\alpha E(t)}{\D t^\alpha}.
\label{eq:ColeColeTime}
\end{equation}

The current response to a voltage step, $U(\omega) = (\J \omega)^{-1}$, when the impedance is $Z = (\J \omega C)^{-1}$ is found as:
\begin{equation}
 I_{step}(\omega)  = \frac{U(\omega)}{Z(\omega)} = \frac{\varepsilon_0 A}{d} \, \varepsilon_r(\omega)  = \frac{\varepsilon_0 A}{d} \,\mathcal{F} \left\{\phi(t) \right\},
\end{equation}
Therefore the current response of the Cole-Cole model is found by inverse Fourier transformation of \eqref{eq:ColeCole} to be \cite{garrappa2016models}:
\begin{equation}
\phi(t) = \frac{1}{\tau} \left(t/\tau \right)^{\alpha-1} \mathrm{E}_{\alpha,\alpha} \left(-\left(t/\tau \right)^\alpha \right),
\end{equation}
where $\mathrm{E}_{\alpha,\beta}$ is the two parameter Mittag-Leffler function: 
\begin{equation}
\mathrm{E}_{\alpha,\beta}(t) = \sum_{n=0}^\infty \frac{t^n}{\mathrm{\Gamma}(\alpha n + \beta)}, \enspace 0<\alpha \le 1.
\label{eq:Mittag-Leffler-2}
\end{equation}
A value  $\beta=1$ results in the standard Mittag-Leffler function $\text{E}_{\alpha}(t) = \text{E}_{\alpha,1}(t)$. Further $\text{E}_{1}(t)$ is the exponential function, which the Mittag-Leffler function generalizes.
The asymptotes of the current response are
\begin{equation}
\phi(t)  \sim
\begin{cases}
  \frac{1}{\tau \mathrm{\Gamma}(\alpha)}  (t/\tau)^{\alpha-1}, \enspace & t \ll \tau \\ 
  \frac{1}{\tau \mathrm{\Gamma}(-\alpha)}  (t/\tau)^{-\alpha-1}, \enspace & t \gg \tau.
\end{cases}
\end{equation}
One will recognize the Curie-von Schweidler law of \eqref{eq:CurieVonSchweidler} as the asymptote for small time arguments as pointed out in  
\cite{de1985memory, holm2020time}.

Since current is proportional to the time derivative of charge, the charge response to a voltage step will be:
\begin{equation}
Q_{step}(\omega) = \frac{I_{step}(\omega)}{\J \omega A} =   \frac{\varepsilon_0}{d}\, \frac{\varepsilon_r(\omega)}{\J \omega} 
= \frac{\varepsilon_0}{d} \,\mathcal{F} \left\{\Psi(t) \right\}
\label{ChargeStepResponse}
\end{equation}
Based on this, the charge response of the Cole-Cole as well as its asymptotes can be found \cite{garrappa2016models}:
\begin{equation}
\Psi(t)  = \mathrm{E}_{\alpha} \left(-\left(t/\tau \right)^\alpha \right) \sim
\begin{cases}
  \exp{[\frac{-(t/\tau)^\alpha}{\mathrm{\Gamma}(\alpha+1)}]}, \enspace & t \ll \tau \\ 
  \frac{(t/\tau)^{-\alpha}}{\mathrm{\Gamma}(1-\alpha)}, \enspace & t \gg \tau.
\end{cases}
\end{equation}
For small time, this is the stretched exponential of \eqref{eq:KWW}.

This demonstrates that the fractional framework makes it possible to show that the two classical temporal responses of  \eqref{eq:CurieVonSchweidler} and \eqref{eq:KWW} correspond to the independently proposed frequency domain Cole-Cole law of \eqref{eq:ColeCole}.

However, there is some ambiguity in these results as the same asymptotic current and  charge responses result from the  more general Havriliak-Negami model. This model is used to model e.g.~molecular chains and polymers:
\begin{equation}\index{Havriliak-Negami model}
{\varepsilon_r}(\omega) = \varepsilon_\infty + \frac{\varepsilon_s - \varepsilon_\infty}{(1 + (\I \omega \tau)^{\alpha})^{\beta}}, 
\quad 0 < \alpha \le 1, \quad 0< \beta < 1,
\label{eq:HavriliakNegami}
\end{equation}
where  $\beta=1$ corresponds to the Cole-Cole model and  $\alpha=1$ is the Cole-Davidson model. As shown in \cite{garrappa2016models}, the Curie-von Schweidler law and the Kohlrausch-Williams-Watt function fit the asymptotes of all these models. 

It should also be noted that the frequency domain expression of \eqref{eqn:CD-factor} corresponds to the Cole-Davidson model ($\beta=-\gamma$ and $\alpha=1$. The 
fractional pseudo-differential operator is therefore central for finding the time-domain relation between $D(t)$ and $E(t)$ for the Cole-Davidson dielectric as was done for the Debye and Cole-Cole models in \eqref{eq:DebyeTime} and \eqref{eq:ColeColeTime} respectively.

\section{Conclusion}

Power laws play an important role in many fields. In linear viscoelasticity and acoustics, there are power-law relaxation responses and attenuation that varies with a frequency power law. This can be modeled with a fractional damper. 

The thickness of a viscous boundary layer is inverse proportional to the square root of frequency or a half-order fractional derivative.  In a poroviscoelastic medium, this is expressed as a viscodynamic operator with a half-order fractional pseudo-differential operator in  Biot theory. 

Observations of power-law and stretched exponential current or charge responses in non-ideal capacitors are compatible with among others the Cole-Cole permittivity model.

A  challenge is to uncover what  it is in the underlying physics of these complex media models that give rise to power-law behavior. There seems to be a specific distribution of multiple relaxation processes which gives rise to the specific characteristics of these media \cite{Nasholm2011}, and it would be very interesting to find out to what extent this property can be derived from first principles.

\end{document}